# Thermal Stress Analysis of the LNG Corrugated Cryogenic Hose During Gas Pre-Cooling Process

*Liu Miaoer[1]   Li Fangqiu[1]   Cheng Hao[1]   Li Endao[1]   Yan Jun[2]   Lu Hailong[2]   Bu Yufeng[2]   Tang Tingting[2]   Lu Zhaokuan[3,*]*

[1] Research and Development Center, CNOOC Gas & Power Group Co. Ltd., Beijing, China

[2] State Key Laboratory of Structural Analysis for Industrial Equipment, School of Mechanics and Aerospace Engineering, Dalian University of Technology, Dalian, China

[3] Ningbo Institute of Dalian University of Technology, Ningbo, China

## ABSTRACT

In this study, thermal-fluid-solid coupled simulations on the gas-phase pre-cooling operation of the corrugated cryogenic hoses were performed. Attention was focused on the temporal evolution and spatial distribution of transient thermal stress in the hose structure caused by convective heat transfer of the cooling medium, Liquefied Natural Gas Boil-Off Gas (BOG). The effects of different corrugated hose parameters, i.e., boundary conditions, hose lengths, BOG inlet flow rates, and corrugation shapes (C-type and U-type), on the transient thermal stress behavior were thoroughly assessed. The thermal stress developed at different locations of the corrugated hoses with these parameters is found to be governed by two major factors: the boundary constraint and local temperature gradient. The objective of this study is to offer practical insights for the structural strength design of corrugated cryogenic hoses and effective pre-cooling strategies, aiming to mitigate structural safety risks caused by excessive thermal stress.

## INTRODUCTION

The exploration and development of natural gas resources in deep-sea regions present both significant opportunities and challenges. These resources, predominantly composed of methane, offer a substantial and relatively untapped energy potential. For deep-sea natural gas development, the collected gas is first transported to a nearby floating liquefied natural gas (LNG) production storage and offloading unit (FLNG) before received by a LNG carrier for transportation back to land (Won et al., 2014). Traditionally, LNG transport between the FLNG and the LNG carrier is achieved by rigid discharge arms, which requires side-by-side alignment of the two vessels with minimal relative motion (Eide et al., 2011). However, under harsh environmental conditions, this requirement is quite difficult to satisfy with the current mooring technology. As an alternative to the rigid discharge arms, corrugated cryogenic hoses (as shown in Fig. 1) has been developed to tackle this problem. With insulation to prevent pipeline leakage and vaporization and the ability to float in the water, it allows for a distance of up to 100 meters between the FLNG and LNG carriers, thus enhancing operational safety in harsh offshore conditions. In recent years, corrugated cryogenic hoses have been increasingly used in offshore LNG production and transportation. Most of the relevant studies focused on the overall design (Yang et al., 2017), structural strength analysis (Yan et al., 2022), and flow and heat transfer characteristics of the internal flow (Yang et al., 2022). The safety issues that may be encountered in the practical field application of the cryogenic hose are not fully understood.

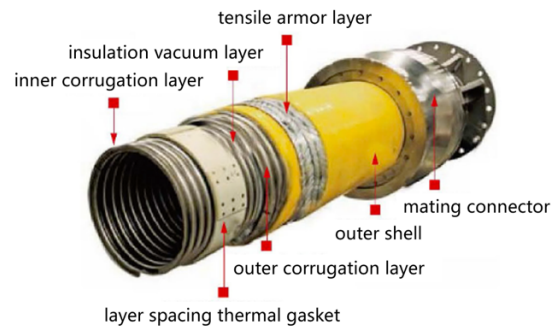

Fig. 1 Diagram of a typical corrugated cryogenic hose

The inner layer of the LNG hose is usually made of austenitic stainless steel (e.g., 316 L), which has a high coefficient of thermal expansion and is therefore susceptible to excessive thermal stresses under severe temperature changes. As a result, the hose needs to be slowly cooled to near its operating temperature (i.e., -162 °C) before transporting LNG to avoid plastic deformation of the structure due to thermal stress or thermal fatigue under repeated operations. However, there has been no widely accepted practice for pre-cooling the corrugated cryogenic hose. In contrast, pre-cooling techniques of the LNG transfer lines (smooth cryogenic pipelines) at LNG receiving terminals is more mature. The pipeline is usually pre-cooled to around -120 °C (gas-phase pre-cooling) by low-temperature nitrogen or LNG boil-off gas (BOG), and then further cooled to its boiling point with LNG (liquid-phase pre-cooling). Although nitrogen is safer than BOG, it has a lower cooling capacity (constant pressure specific heat capacity is 50% of BOG). The nitrogen in the pipeline during the pre-cooling process is prone to stratification, resulting in a temperature difference of more than 50 °C between the top

*Corresponding Author: Zhaokuan Lu (email: luzk_nbi@dlut.edu.cn)

and bottom of the pipeline, which exposes the pipeline to excessive localized thermal stresses and the risk of tearing at the weld (Chunsheng et al., 2020).

Since the density of BOG is less sensitive to temperature variation, there are relatively few working conditions in which the temperature difference between the upper and lower surfaces of the pipeline is large due to the stratification of cryogenic gas. Therefore, BOG has become a mainstream pre-cooling medium in the gas-phase pre-cooling stage of the LNG discharge lines (Yuan et al., 2015).

Research related to the pre-cooling of LNG pipes has also been predominantly focusing on the smooth cryogenic pipelines (up to several kilometers) rather than the corrugated cryogenic hoses, and most of these studies used Computational Fluid Dynamics (CFD) approach to analyze the flow and heat transfer in the pipe. One-dimensional CFD model has been employed to simulate the BOG pre-cooling of LNG pipelines and assess the temperature variation of the pipe structure under different pre-cooling strategies, e.g., BOG flow rate (Lu et al., 2012). Zhang et al. (2017) simulated the single-phase gas flow and pipe temperature changes during the BOG pre-cooling with a three-dimensional CFD model. Similar to high-temperature steam pipelines, long-distance LNG pipelines are usually equipped with expansion bends at regular intervals to absorb the thermal expansion, among which $\pi$-type expansion bend is most used. Cai (2021) performed CFD simulation of BOG pre-cooling of large diameter pipeline with expansion bend. The structural temperature distribution and heat transfer characteristics of the low-temperature gas passing through the expansion bend were analyzed. The results showed that the wall temperature increases along the direction of BOG flow and decreases at the bend. Wang et al. (2020) performed a similar simulation for both gas-phase and liquid-phase pre-cooling for a large-diameter pipeline with an expansion bend. They used BOG to first lower the pipe to -120°C and then used LNG with small flow rate to further lower the temperature to -150°C. Song (2022) obtained the thermal stress of a section of pipeline with expansion bends. He used the temperature monitoring data of the discharge pipeline at the LNG receiving station as thermal load and then calculated the stress and displacement changes of the expansion bends under different constraints. Chen (2019) used the Species Transport Model (STM) and Volume of Fluid (VOF) model to simulate the temperature distribution and the structural thermal stresses caused by temperature loads in a long-distance LNG receiving terminal pipeline during the gas-phase precooling and liquid-phase precooling phases, respectively. Overall, the simulation of flow and heat transfer characteristics in the long-distance smooth LNG pipelines using the CFD method is still challenging due to the large number of degree-of-freedom required for the computation.

Compared with the long-distance LNG pipelines with smooth inner walls, there are fewer studies on the pre-cooling of corrugated cryogenic hoses. The existing studies mainly focused on the temperature variation of the corrugated structure during the precooling process (Wu et al., 2024). The authors (Liu et al., 2023) conducted a CFD simulation on the LNG pre-cooling of the corrugated cryogenic hoses and found that the LNG inlet flow rate has a significant effect on the temperature gradient of the corrugated structure. Although the corrugation can reduce the thermal stress to a certain extent, the temperature gradient caused by drastic temperature change during the pre-cooling process may still lead to excessive local thermal stress, which causes structural failure. Therefore, it is necessary to analyze the characteristics of thermal stress developed during the pre-cooling process of the corrugated cryogenic hose. In this study, the thermo-fluid-solid coupling simulation of the corrugated cryogenic hose during the BOG gas-phase pre-cooling is carried out. Distribution and evolution of the transient thermal stress are analyzed in-depth. Parametric studies on the effects of pre-cooling strategies and hose geometric design are performed. The results and implications may provide references for pre-cooling field operations and thermal stress resistant design of the corrugated cryogenic hose.

## COMPUTATIONAL MODEL SETUP

In this study, a coupled thermal-fluid-solid model is constructed by coupling FLUENT and Static Structural modules in ANSYS Workbench (Ansys, 2020) to simulate the turbulent gas mixture inside the hose, the conjugate heat transfer between fluid and solid and the thermal stress of the corrugated hose. The model geometry and governing equations of the numerical model are described in the following subsections.

### Geometric Model Setup

Since more than 90% of the natural gas recovered from the gas well is methane and the other components each constitutes a small portion, it is assumed that the BOG used for pre-cooling is pure methane gas. Given that the density of -120 °C methane gas is 1.277 kg/m³ at atmospheric pressure, similar to the density of air, it is reasonable to postulate that buoyancy does not play a significant role during the convection of BOG in corrugated hose initially filled with air. Therefore, the gravity term is neglected in the model, and the hose can therefore be simplified to a two-dimensional axisymmetric structure for efficient calculation. Since the design of corrugated cryogenic hose is quite diversified, including corrugation layer, insulation layer, armor layer, among other functional layers. Without loss of generality, this paper only models the corrugation layer, which is the common innermost layer for all the existing corrugated cryogenic hoses. The geometric parameters of the corrugation layer are defined as shown in Fig. 2, where $R$ is the hose radius measured at the narrowest part of the corrugation, $r$ is the corrugation radius, $h$ is the depth of corrugation, and $\lambda$ is the corrugation wavelength. In this study, we set $R = 8.55$ cm, $r = 0.635$ cm, $\lambda = 2.54$ cm, and set the corrugation layer wall thickness to be 0.5 mm.

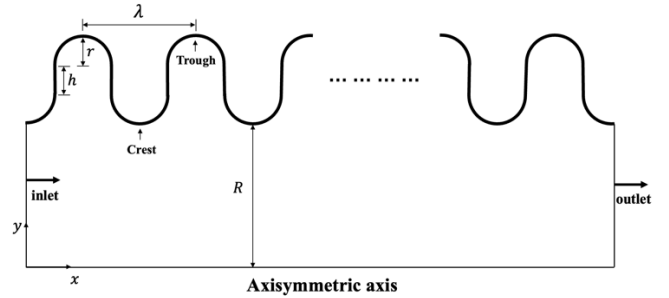

Fig. 2 Two-dimensional axisymmetric corrugated hose model

### Turbulence Modeling

The flow inside the corrugated hose is assumed to be incompressible turbulent flow since the velocity of the mixed flow of BOG and air inside the hose is much lower than the speed of sound. To describe the transient dynamics of this flow, the mass conservation and momentum conservation equations are constructed using the unsteady Reynolds averaging (URANS) approach:

$$\frac{\partial \rho}{\partial t} + \nabla \cdot \boldsymbol{u_f} = 0 \quad (1)$$

$$\frac{\partial}{\partial t}(\rho \boldsymbol{u_f}) + \rho(\boldsymbol{u_f} \cdot \nabla)\boldsymbol{u_f}$$
$$= -\nabla\left(P + \frac{2}{3}\rho k\right) + \nabla \cdot (\mu + \mu_T)\left(\nabla \boldsymbol{u_f} + (\nabla \boldsymbol{u_f})^T\right) \quad (2)$$

, where $\rho$ is the fluid density, $t$ is the time, $\boldsymbol{u}_f$ is the flow velocity vector, $P$ is the fluid pressure, and $\mu$ is the fluid viscosity. $k$ is the turbulent kinetic energy, and $\mu_T$ is the eddy viscosity, both generated by the Boussinesq approximation to capture the effects of turbulent flow oscillations. In this paper, we use the standard $k-\epsilon$ closure model (Launder and Spalding, 1972) that expresses the turbulent viscosity in terms of the turbulent kinetic energy $k$ and its dissipation rate $\epsilon$:

$$\mu_t = \rho C_\mu \frac{k^2}{\epsilon} \qquad (3)$$

, where $C_\mu$ is an empirical constant.

In terms of the boundary conditions of the fluid domain (boundaries defined as in Fig. 2), normal BOG flow and zero gauge pressure are imposed at the model inlet and outlet, and axisymmetric boundary condition is set at the center of the pipe. Non-slip boundary condition is prescribed at the wall, and the enhanced wall treatment is employed to improve the near-wall behavior of the standard $k-\epsilon$ model. Quadrilateral-dominant method is used for fluid domain meshing, and it is ensured that the dimensionless wall distance $y^+$ of the first layer of the mesh near the wall is approximately equal to 1.

***Turbulence model validation.*** To verify the accuracy of the standard $k-\epsilon$ model with enhanced wall treatment in the 2D axisymmetric configuration, the flow velocity profiles inside the corrugated hose obtained by an existing experiment (Calomino et al., 2015) are compared with those calculated by numerical simulation with the same geometric parameters (same $R$ and $\lambda$ values with the model in Fig. 2 and $h = 0$, $r = 0.6$ cm). The inlet flow velocity is 0.44 m/s. The comparison of axial flow velocity ($U_x$) at the crest and trough of the corrugation (defined in Fig. 2) along the radial direction (y direction) is shown in Fig. 3. It can be seen that the two-dimensional axisymmetric turbulence model used in this study can well capture the flow velocity distribution in the corrugated hose, including the negative flow velocity near the wall due to the boundary layer separation, see Fig. 3 (a).

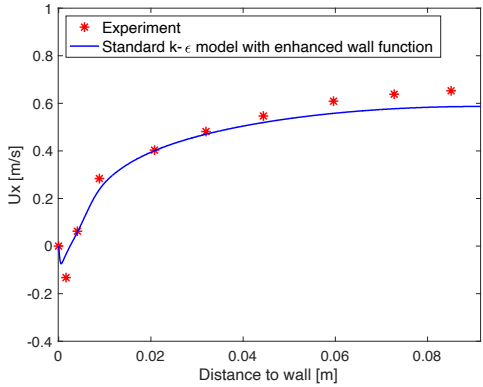

(a) Crest

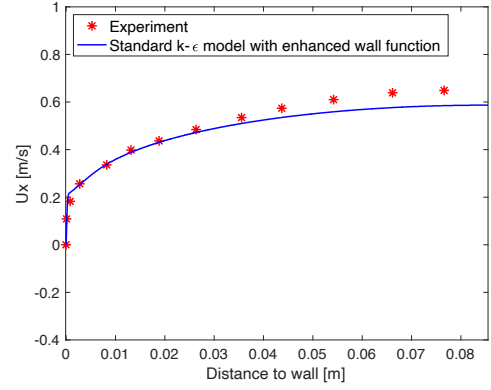

(b) Trough

Fig. 3 Comparison between the experimental and numerical flow velocity distribution

## Species Transport Modeling

During the pre-cooling process, flow mixing occurs between the BOG and the initial air resides inside the hose. The convective and diffusive motions in the gas mixing process can be described using the Species Transport Model (STM). The specific form of the transport equations for each component $i$ is written in the following form:

$$\frac{\partial}{\partial t}(\rho Y_i) + \nabla \cdot (\rho \boldsymbol{u}_f Y_i) = -\nabla \cdot \vec{J}_i + R_i + S_i \qquad (4)$$

, where $Y_i$ is the mass fraction of each component. The first term on the left-hand side (LHS) of the equation is the transient variation of the mass fraction of each component. The second term is the convective term for the component mass fraction due to flow transport. $\boldsymbol{u}_f$ is the flow velocity vector, through which Eq. 4 is coupled with the mass conservation and momentum equations (Eq. 1 and 2). $\vec{J}_i$ on the right-hand side (RHS) of Eq. 4 is the component mass diffusion term, and $R_i$ is the net rate of production of each component by chemical reaction (since BOG and air do not involve chemical reactions, $R_i$ is set to 0). $S_i$ is the source term for each component (the inlet BOG input is realized by this term).

## Heat Transfer Modeling

To simulate the heat exchange within and between the flow mixture and the hose structure, energy conservation equations for the fluid and the solid are solved and the two are coupled by conjugate heat transfer constraints. The governing equation describing the fluid convective heat exchange is:

$$\frac{\partial}{\partial t}(\rho E) + \nabla \cdot (\boldsymbol{u}_f(\rho E + p)) = \nabla \cdot \left(k_{eff}\nabla T - \sum_j h_j \vec{J}_j + (\bar{\bar{\tau}}_{eff} \cdot \boldsymbol{u}_f)\right) \qquad (5)$$

The LHS of Eq. 5 represents the energy transfer due to convective heat transfer. The first term on the RHS of the equation represents the energy transfer due to thermal conduction. Effective conductivity $k_{eff}$ depends on the thermal conductivity of the fluid and turbulent thermal conductivity. The second and third terms on the RHS represent the energy transfer due to gas mixing and viscous dissipation.

The energy conservation equation of the hose structure has a simpler

form since turbulence is not involved:

$$\frac{\partial}{\partial t}(\rho h) + \nabla \cdot (\boldsymbol{u_s}\rho_s h) = \nabla \cdot (k_s \nabla T) \quad (6)$$

Eq. 6 regulates the energy balance between the thermal convection (LHS) due to the movement of the structure ($\boldsymbol{u_s}$ is the velocity vector of the structure) and the thermal conduction inside the structure (RHS). $k_s$ is the thermal conductivity of the structure, and the sensible enthalpy $h$ can be calculated by integrating the specific heat capacity from the reference temperature to the current temperature.

For the boundary conditions, the temperature at the fluid domain inlet is set to be -120 °C. The hose left and right ends and outer surface are set to be zero heat flux boundary (assuming no heat exchange between the hose and external environment). Conjugate heat transfer constraints are imposed at the fluid-solid interface to allow heat transfer between the gas mixture and the hose structure:

$$T_s^w = T_f^w \quad (7)$$

$$-k_s \frac{\partial T_s^w}{\partial n_s} = k_f \frac{\partial T_f^w}{\partial n_f} \quad (8)$$

where $T_s^w$ and $T_f^w$ are the temperature of the hose and the fluid at their interface, $n_s$ and $n_f$ are the normal direction at the interface, $k_f$ is the thermal conductivity of the fluid, respectively. Eq. 7 and Eq. 8 ensure the continuity of temperature and conductive heat flux at the fluid-solid interface.

**Structural Response Modeling**

Governing equation of the structural response is based on Newton's second law in the Lagrangian framework:

$$\nabla \cdot \boldsymbol{\sigma_s} + \boldsymbol{F} = 0 \quad (9)$$

where $\boldsymbol{\sigma_s}$ is the total stress vector, which can be obtained from the total strain vector $\boldsymbol{\epsilon_s}$ and the elastic stiffness matrix $D$ as $\boldsymbol{\sigma_s} = D\boldsymbol{\epsilon_s}$. $\boldsymbol{F}$ is the structural body force vector. Since the structural response in the pre-cooling process is mainly induced by thermal load, the structural gravity is ignored, and $\boldsymbol{F}$ is set to 0. Because the structure is not subject to external force, the total strain vector becomes the thermal strain $\boldsymbol{\epsilon}^{th}$ which can be expressed as:

$$\boldsymbol{\epsilon}^{th} = (T - T_{ref})\boldsymbol{\alpha_{sc}} \quad (10)$$

, where $T$ and $T_{ref}$ are the structure temperature and the reference temperature (i.e., the temperature at which the thermal strain of the material is zero), and $\boldsymbol{\alpha_{sc}}$ is the vector of secant thermal expansion coefficient. It is assumed that the material of the hose structure is isotropic and the $\boldsymbol{\alpha_{sc}}$ becomes a constant. A bilinear isotropic hardening model is used to handle potential plastic deformation. For spatial discretization, the structural domain is divided by all-quadrilateral element of uniform density.

**Material Property Settings**

As mentioned earlier, the inlet BOG was set to be 100% methane and the initial air in the hose is composed of 21% oxygen and 79% nitrogen. Each gas component uses the temperature-dependent material properties intrinsically defined in FLUENT. The material of the corrugated hose is 316 L stainless steel whose material properties are set based on the data from the National Institute of Standards and Technology (NIST, 2023). The reference temperature $T_{ref}$ is set to be 19.8 °C. To account for plastic deformation, the yield stress of the material is defined as 225 MPa and the tangent modulus as 2091 MPa.

**Thermal-fluid-solid Coupling Approach**

As the stress response of the hose during the pre-cooling process is predominantly influenced by temperature variation, a common method used in the study of thermal-fluid-solid coupling in cryogenic equipment (Kim et al., 2018; Zhang et al., 2020) involves mapping the transient structural temperature field, obtained from the flow heat transfer and fluid-solid conjugate heat transfer calculations in FLUENT, onto the structural model in the steady-state finite element calculation module, Static Structure. This temperature field is then used as a load to calculate the corresponding structural thermal stress at each time step.

RESULTS AND DISCUSSION

In order to provide a reference for the design of corrugated cryogenic hose and pre-cooling strategy, this section evaluates the temporal and spatial variation of the thermal stress developed in the corrugated hose structure with different overall lengths, pre-cooling flow velocities, and corrugation shapes. To also consider the effects of boundary constraints, clamped-clamped and clamped-free (inlet end clamped and outlet end free) boundary conditions are applied to the structural model for each simulation condition. The total simulation time is set to be 1800 s to make sure the simulation covers the entire gas-phase pre-cooling process (temperature down to around -120 °C). Time history of the equivalent stress at the inner surfaces of three consecutive crests and troughs near the hose joint are obtained to reflect the spatial variations of the thermal stresses (the sampling points are defined in Fig. 4). Grid independent analysis for each simulation case suggests that the thermal stress at the sampled points is converged when the element size is around 5e-4 m.

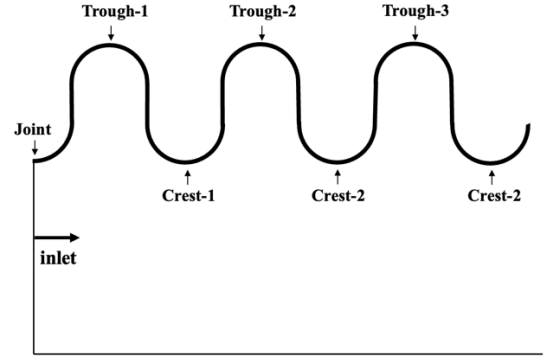

Fig. 4 Sampled location for thermal stress

**Effects of Hose Length on Thermal Stress**

To investigate the effects of hose length on the thermal stress generated during the precooling process, a corrugated hose model with C-type corrugation (h = 0) is built. The hose length is set to be 10, 20, and 40 wavelengths, and the inlet flow velocity is 0.4 m/s. Transient thermal stress of the three hoses under clamped-clamped and clamped-free boundary conditions are shown in Fig. 5 ~ Fig. 7. It can be seen that the temporal variations of the thermal stress for hoses with different lengths show a similar trend.

In case of clamped-clamped boundary condition, the thermal stress rapidly reach a maximum and stabilize at the joint and trough-1 location closest to it due to the local stress concentration generated by thermal contraction and the fixed constraint of the clamped end. The thermal stress experienced at distant points from the joint (i.e., crest-1, trough-2, crest-2, trough-3, crest-3) initially peaks at a local maximum, subsequently escalating towards a steady-state peak. This peak stress progressively diminishes as the distance from the joint increases. The magnitude and timing of the maximum thermal stress at each location are shown in Table 1. As the hose length increases, the maximum thermal stress at the joint remains almost the same while that at the other locations tends to grow. This may be attributed to the fact that the hose becomes more flexible as its length increases, resulting in greater degree of thermal contraction under the same temperature variation, and thus generating stronger thermal stress in the regions that are less affected by the boundary constraint. Overall, the maximum thermal stress for all structural locations happens in the steady state, i.e., after the pre-cooling is completed. Therefore, instead of transient simulations, steady-state structural analyses with thermal loads equal to the temperature of the inlet cryogenic gas can be conducted to determine the maximum thermal stress of clamped-clamped C-type corrugated hoses, which is much more computationally efficient.

Under the clamped-free boundary condition, the maximum thermal stress at all assessed locations except the joint and trough-1 occurs during the pre-cooling process rather than the end of it (the magnitude and timing of the maximum thermal stress are shown in Table 2). This is because the thermal stress at locations far from the joints is no longer dominated by the fixed constraint but by the transient temperature gradient generated by the passage of the BOG. As the temperature becomes more evenly distributed as the cooling continues, the thermal stress gradually diminishes after the transient maxima. As a result, transient analysis has to be performed in order to determine the maximum thermal stress at the locations away from the clamped end for C-type corrugated hoses under clamped-free boundary conditions. Also, the magnitude of the maximum thermal stress at each assessed location is essentially independent of the hose length. Apart from the joint and trough-1 area, where the maximum thermal stress is consistent with that of the clamped-clamped hose due to the influence of fixed constraint, the maximum thermal stress at all other locations is less than that observed in the case of clamped-clamped boundary condition.

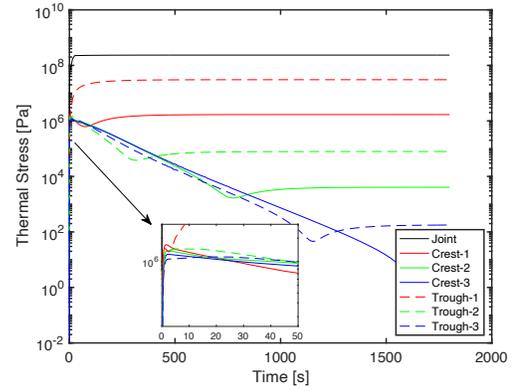

(b) clamped-free boundary condition

Fig. 5 Thermal stress history of the hose with 10 wavelengths (flow rate is 0.4 m/s)

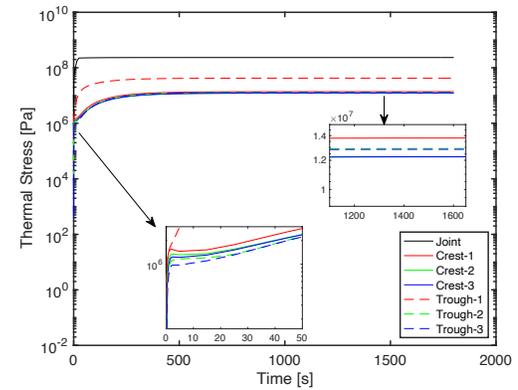

(a) clamped-clamped boundary condition

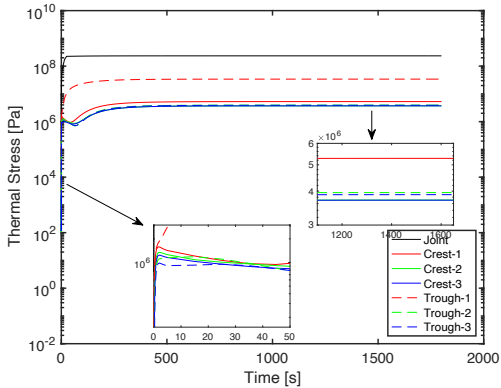

(a) clamped-clamped boundary condition

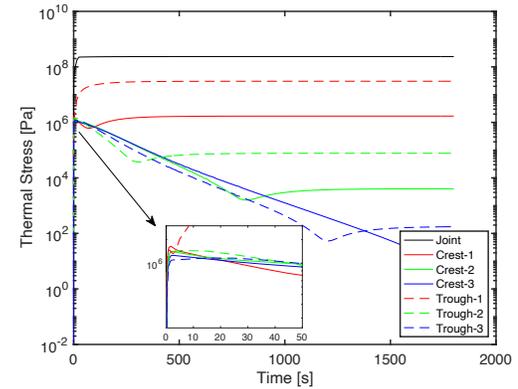

(b) clamped-free boundary condition

Fig. 6 Thermal stress history of the hose with 20 wavelengths (flow rate is 0.4 m/s)

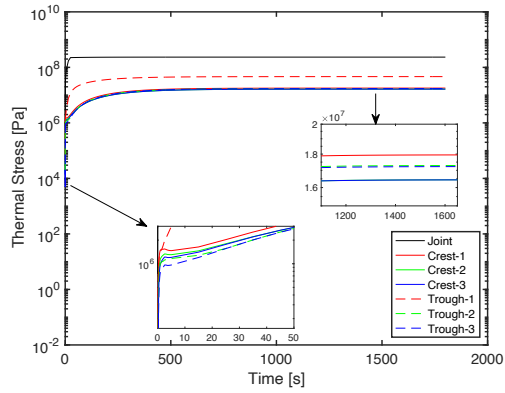

(a) clamped-clamped boundary condition

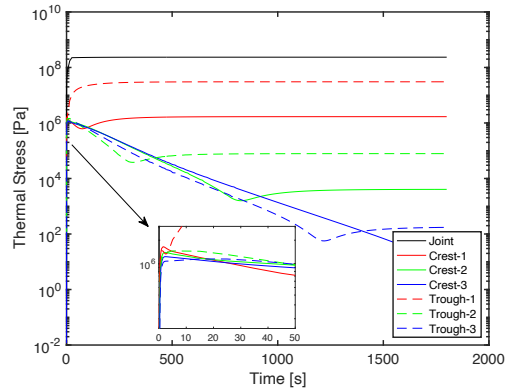

(b) clamped-free boundary condition

Fig. 7 Thermal stress history of the hose with 40 wavelengths (flow rate is 0.4 m/s)

Table 1 The value (Pa) and timing of max thermal stress at the sampled locations (clamped-clamped boundary condition)

| Location | 10 wavelengths | 20 wavelengths | 40 wavelengths |
| --- | --- | --- | --- |
| Joint | 2.35e8 (steady state) | 2.34e8 (steady state) | 2.34e8 (steady state) |
| Trough-1 | 3.40e7 (steady state) | 4.24e7 (steady state) | 4.65e7 (steady state) |
| Crest-1 | 5.28e6 (steady state) | 1.38e7 (steady state) | 1.8e7 (steady state) |
| Trough-2 | 3.95e6 (steady state) | 1.29e7 (steady state) | 1.73e7 (steady state) |
| Crest-2 | 3.71e6 (steady state) | 1.23e7 (steady state) | 1.64e7 (steady state) |
| Trough-3 | 3.88e6 (steady state) | 1.28e7 (steady state) | 1.72e7 (steady state) |
| Crest-3 | 3.71e6 (steady state) | 1.23e7 (steady state) | 1.64e7 (steady state) |

Table 2 The value (Pa) and timing of max thermal stress at the sampled locations (clamped-free boundary condition)

| Location | 10 wavelengths | 20 wavelengths | 40 wavelengths |
| --- | --- | --- | --- |
| Joint | 2.35e8 (steady state) | 2.35e8 (steady state) | 2.35e8 (steady state) |
| Trough-1 | 3.04e7 (steady state) | 3.04e7 (steady state) | 3.04e7 (steady state) |
| Crest-1 | 1.69e6 (steady state) | 1.69e6 (steady state) | 1.69e6 (steady state) |
| Trough-2 | 1.46e5 (5 s) | 1.45e6 (5 s) | 1.46e6 (5 s) |
| Crest-2 | 1.40e6 (2) | 1.40e6 (2 s) | 1.40e6 (2 s) |
| Trough-3 | 1.15e6 (15 s) | 1.17e6 (15 s) | 1.15e6 (15 s) |
| Crest-3 | 1.24e6 (2.5 s) | 1.26e6 (2.5 s) | 1.23e6 (2.5 s) |

### Effects of BOG Flow Velocity on Thermal Stress

To study the effects of BOG flow velocity on the thermal stress in hose structure during the pre-cooling process, the C-type corrugated hose model with 20 wavelengths is used. The structural responses under inlet flow velocities of 0.1 m/s, 0.2 m/s, 0.4 m/s, and 0.8 m/s are simulated, with the results of transient thermal stress at various sampling points presented in Fig. 6, 8~10.

The simulation results indicate that under the clamped-clamped boundary condition, the maximum thermal stress at all points occurs in the steady state. The higher the flow velocity, the sooner the thermal stress reaches this steady-state maxima. Since the maximum thermal stress appears in the steady state, the maximum thermal stress of the structure does not depend on the flow velocity but only on the temperature of the pre-cooling medium and the properties of the hose itself. Therefore, for the C-type corrugated hose with both ends clamped, the maximum thermal stress during the BOG pre-cooling process is independent of the flow velocity.

Under the clamped-free boundary condition, the maximum thermal stress near the joint (i.e., at the joint and trough-1 location) still occurs after the pre-cooling ends. However, the maximum thermal stress at points further away from the joint occurs during the passage of the BOG, rather than in the steady state, and decreases progressively along the length of the hose. As shown in Table 3, which lists the magnitude and timing of the maximum thermal stress, the greater the flow velocity, the larger the maximum stress at points away from the joint, and the earlier it appears. The relationship between the maximum thermal stress and flow velocity is depicted in Fig. 11. This is primarily because higher BOG flow rates enhance convective heat transfer, leading to a faster local temperature drop and generating larger local temperature gradients. As a result, greater thermal stress is induced in areas far from the joint, where the magnitude of the thermal stress is governed by local temperature gradient.

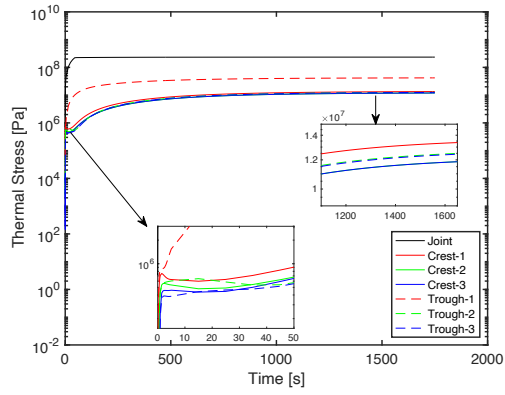

(a) clamped-clamped boundary condition

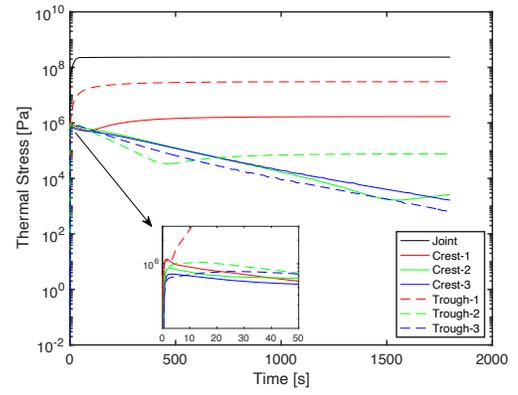

(b) clamped-free boundary condition

Fig. 9 Thermal stress history when BOG flow rate equals to 0.2 m/s

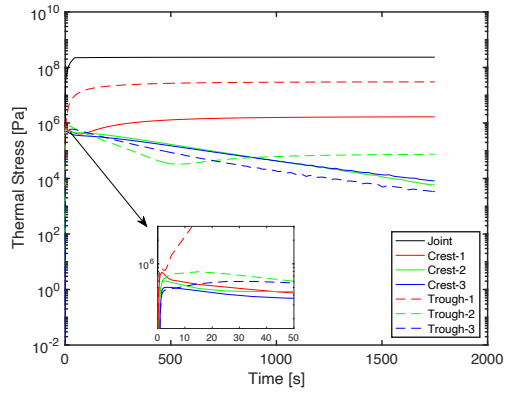

(b) clamped-free boundary condition

Fig. 8 Thermal stress history when BOG flow rate equals to 0.1 m/s

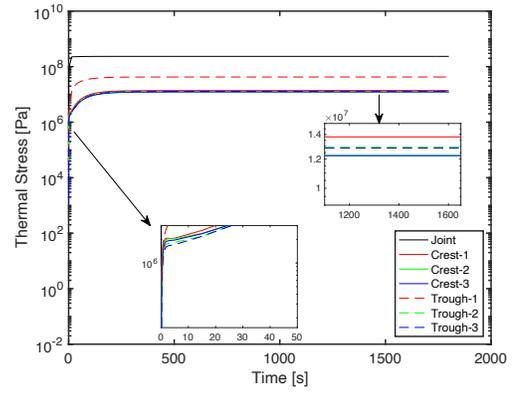

(a) clamped-clamped boundary condition

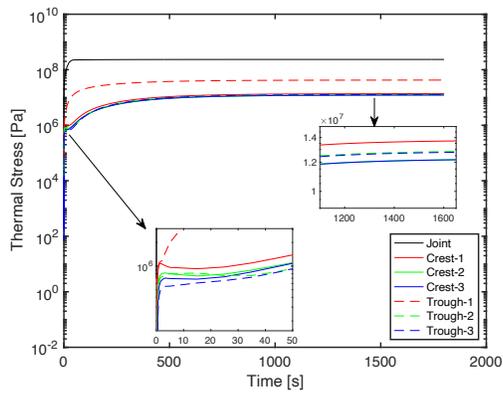

(a) clamped-clamped boundary condition

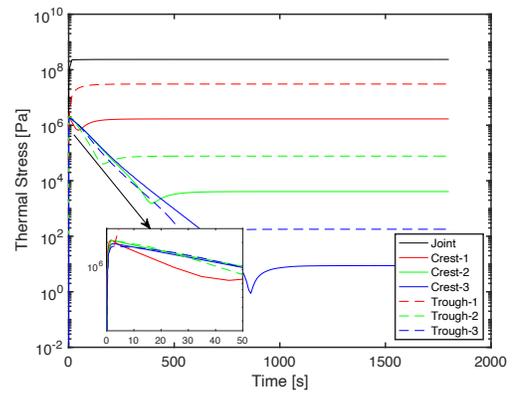

(b) clamped-free boundary condition

Fig. 10 Thermal stress history when BOG flow rate equals to 0.8 m/s

Table 3 The value (Pa) and timing of max thermal stress at the sampled locations (clamped-free boundary condition)

| Location | 0.1 m/s | 0.2 m/s | 0.4 m/s | 0.8 m/s |
|---|---|---|---|---|
| Joint | 2.35e8 (steady state) | 2.35e8 (steady state) | 2.35e8 (steady state) | 2.35e8 (steady state) |
| Trough-1 | 3.04e7 (steady state) | 3.04e7 (steady state) | 3.04e7 (steady state) | 3.04e7 (steady state) |
| Crest-1 | 1.67e6 (steady state) | 1.67e6 (steady state) | 1.67e6 (steady state) | 2.14e6 (1.5 s) |
| Trough-2 | 7.93e5 (15 s) | 1.04e6 (15 s) | 1.45e6 (5 s) | 2.05e6 (4.5 s) |
| Crest-2 | 6.05e5 (2.5 s) | 8.81e5 (2.5 s) | 1.41e6 (2.5 s) | 2.10e6 (2.5 s) |
| Trough-3 | 5.96e5 (35 s) | 7.98e5 (25 s) | 1.17e6 (15 s) | 1.79e6 (3 s) |
| Crest-3 | 5.01e5 (3.5 s) | 7.39e5 (3 s) | 1.26e6 (2.5 s) | 1.92e6 (2.5 s) |

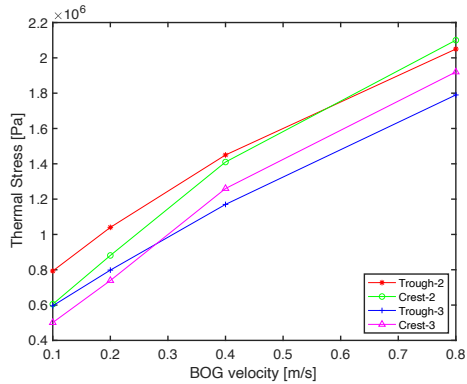

Fig. 11 The relationship between max thermal stress and BOG flow velocity at different locations

### Effects of Corrugation Shape on Thermal Stress

To assess the effects of corrugated hose design on thermal stress during the pre-cooling process, U-type corrugated hoses with corrugation depths of $h = 0.5$ cm and $h = 1$ cm are constructed based on the C-type corrugated hose used in the previous subsections. A BOG inlet flow velocity of 0.4 m/s is set for these tests. The transient thermal stress for these two types of corrugated hoses under both clamped-clamped and clamped-free boundary conditions are shown in Fig. 12 and 13. The maximum thermal stress magnitude and its timing for C-type and U-type corrugated hoses are listed together in Tables 4 and 5 for comparison.

Under the clamped-clamped condition, the transient thermal stress of the C-type and $h = 0.5$ cm U-type corrugated hoses are similar, with the maximum stress occurring after the pre-cooling completes. The U-type corrugated hose, due to its better flexibility, exhibited lower maximum thermal stresses at all sampling points compared to the C-type. On the other hand, the $h = 1$ cm U-type corrugated hose, with even greater flexibility, experienced its maximum thermal stress at locations away from the joint after the passage of the BOG rather than the end of pre-cooling, with the magnitude lower than that of the C-type and $h = 0.5$ cm U-type hoses. This indicates that the maximum thermal stress could be induced by transient local temperature gradient even under clamped-clamped boundary condition if the corrugated hose is flexible enough.

Under the clamped-free boundary condition, the maximum thermal stress in both U-type corrugated hoses, except at the joint and trough-1 location, occurred after the passage of the BOG. The maximum thermal stress at the joint was the same as that of the C-type hose. Compared to the C-type hose, the deeper corrugation of the U-type hoses reduced the thermal stress at positions mainly affected by constraints (trough-1 location) but increased it in areas dominated by temperature gradients, away from the joint. This is because deeper corrugations make the crests and troughs more prone to uneven heating due to boundary layer separation at the crests as the BOG passes, leading to larger local temperature gradients and thus greater thermal stress.

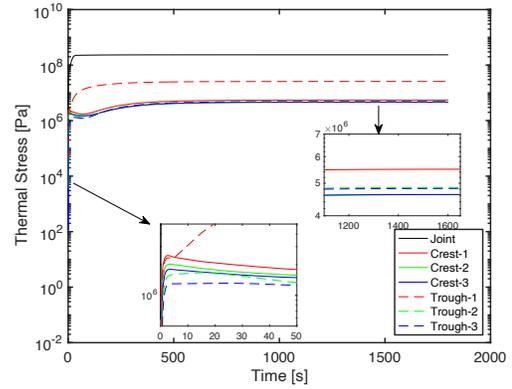

(a) clamped-clamped boundary condition

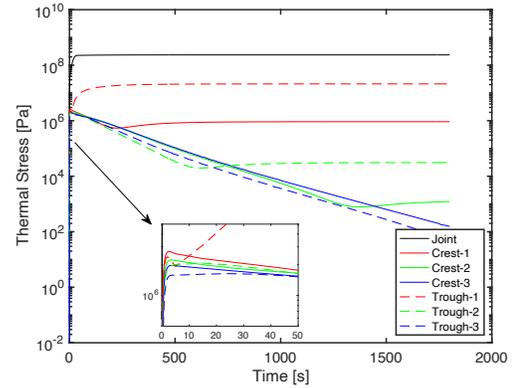

(b) clamped-free boundary condition

Fig. 12 Thermal stress history of the corrugated hose with $h = 0.5$ cm

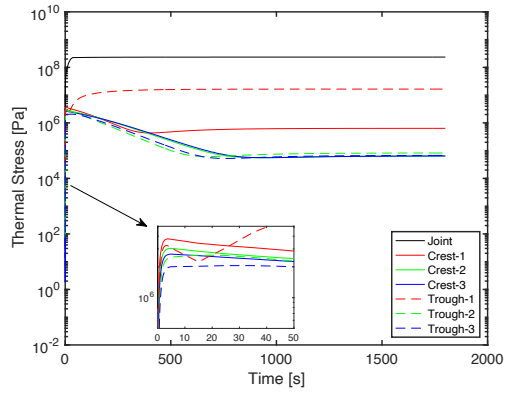

(a) clamped-clamped boundary condition

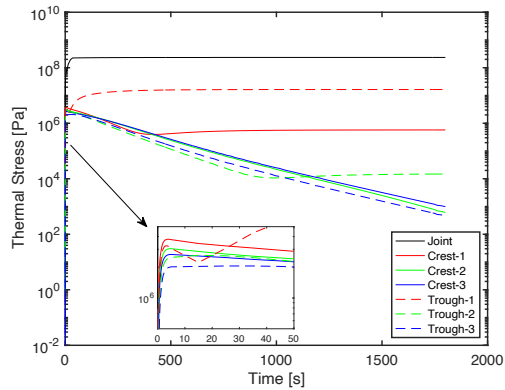

(b) clamped-free boundary condition

Fig. 13 Thermal stress history of the corrugated hose with h = 1 cm

Table 4 The value (Pa) and timing of max thermal stress at the sampled locations (clamped-clamped boundary condition)

| Location | C type | U type (h = 0.5 cm) | U type (h = 1 cm) |
|---|---|---|---|
| Joint | 2.35e8 (steady state) | 2.35e8 (steady state) | 2.35e8 (steady state) |
| Trough-1 | 4.24e7 (steady state) | 2.58e7 (steady state) | 1.65e7 (steady state) |
| Crest-1 | 1.38e7 (steady state) | 5.50e6 (steady state) | 3.76e6 (3.5 s) |
| Trough-2 | 1.29e7 (steady state) | 4.85e6 (steady state) | 2.61e6 (15 s) |
| Crest-2 | 1.23e7 (steady state) | 4.63e6 (steady state) | 3.04e6 (4.5 s) |
| Trough-3 | 1.28e7 (steady state) | 4.82e6 (steady state) | 2.06e6 (25 s) |
| Crest-3 | 1.23e7 (steady state) | 4.63e6 (steady state) | 2.67e6 (4.5 s) |

Table 5 The value (Pa) and timing of max thermal stress at the sampled locations (clamped-free boundary condition)

| Location | C type (h = 0) | U type (h = 0.5 cm) | U type (h = 1 cm) |
|---|---|---|---|
| Joint | 2.35e8 (steady state) | 2.35e8 (steady state) | 2.35e8 (steady state) |
| Trough-1 | 3.04e7 (steady state) | 2.12e7 (steady state) | 1.64e7 (steady state) |
| Crest-1 | 1.67e6 (steady state) | 2.69e6 (2.5 s) | 3.77e6 (4 s) |
| Trough-2 | 1.45e6 (5 s) | 2.07e6 (15 s) | 2.62e6 (15 s) |
| Crest-2 | 1.41e6 (2.5 s) | 2.23e6 (3.5 s) | 3.04e6 (4.5 s) |
| Trough-3 | 1.17e6 (15 s) | 1.64e6 (25 s) | 2.06e6 (25 s) |
| Crest-3 | 1.26e6 (2.5 s) | 1.96e6 (3.5 s) | 2.67e6 (4.5 s) |

CONCLUSION

This paper delves into the structural thermal stress response of corrugated cryogenic hoses using Liquefied Natural Gas Boil-Off Gas (BOG) during the gas-phase pre-cooling stage. We employed Computational Fluid Dynamics (CFD) and Finite Element Method (FEM) simulations to explore the effects of different hose boundary constraints, hose lengths, BOG flow rates, and corrugation shapes on the transient variation and spatial distribution of the structural thermal stress. The main conclusions of this study are as follows:

(1) Under the clamped-clamped boundary condition, the maximum thermal stress in C-type corrugated hoses of varying lengths occurs after the completion of pre-cooling. The maximum thermal stress is highest near the joints and gradually decreases further away. As the hose length increases, the maximum thermal stress near the joint remains constant, but the maximum thermal stress at locations further from the joint increases due to greater thermal strain caused by increased hose flexibility. Under the clamped-free boundary condition, the maximum thermal stress at positions downstream of the hose, other than near the joint, is significantly influenced by local temperature gradients and occurs after the transient passage of the BOG, independent of hose length.

(2) Since the maximum thermal stress of the C-type corrugated hose under the clamped-clamped condition occurs in the steady state after the pre-cooling, the BOG flow rate does not affect the maximum thermal stress. For the hoses with one end clamped and the other free, higher BOG flow rates increase the maximum thermal stress occurring after the BOG passage at locations away from the joint, due to larger temperature gradients generated by enhanced convective heat transfer.

(3) With a positive corrugation depth $h$, the hose transforms from C-type ($h = 0$) to U-type ($h > 0$) with better flexibility. Nevertheless, the overall maximum thermal stress still occurs at the joint position after pre-cooling, regardless of corrugation shape. However, for the U-type hose with $h = 1$ cm, even under the clamped-clamped condition, the maximum thermal stress away from the joint happens during the pre-cooling process, not after it. This indicates that with greater corrugation depth, the thermal stress in downstream areas of the hose is more likely to be governed by local temperature gradients rather than constraints. Under the clamped-free boundary condition, deeper corrugations make it more likely for uneven heating to occur at the crests and troughs of the corrugated structure, leading to greater thermal stress.

(4) Overall, during the BOG gas-phase pre-cooling process, regardless of the boundary condition at the hose outlet end, the maximum thermal stress near the joint always occurs after pre-cooling and can be obtained through steady-state analysis. Although the maximum thermal stress at downstream locations away from the joint is smaller, its magnitude and timing depend on BOG flow rate, hose boundary conditions, and corrugation shape, which requires transient analysis for certain situations.

By analyzing the distribution and evolution of the maximum thermal stress along the hose during the gas-phase pre-cooling process, this study aims to provide insights for pre-cooling strategies and better local strength design of the corrugated hose to prevent unnecessary structural failures due to excessive thermal stress.

ACKNOWLEDGEMENTS

This research was financially supported by the National Natural Science Foundation of China (No. 52201312, 52301315), the National Key R&D Program of China (2021YFC2801602), the Fundamental Research Funds for the Central Universities (DUT22ZD209, DUT22QN251), the Liaoning Province's Xing Liao Talents Program (XLYC2002108), and the Ningbo Key R&D Program (2022Z061, 2023Z050, 2023Z055). These supports are gratefully acknowledged.